      [1999/12/01 v1.4c Il Nuovo Cimento]
\documentclass{cimento}

\title{Observing GRBs with TMT}
\author{David Crampton \atque Luc Simard\from{ins:x}}
\instlist{\inst{ins:x} Herzberg Institute of Astrophysics, NRC Canada, Victoria V9E 2E7  \\
and TMT Project Office, Pasadena, CA 91107, USA}

\PACSes{\PACSit{98.70.Rz}{gamma-ray sources; gamma-ray bursts}
\PACSit{95.55.Cs}{Ground-based ultraviolet, optical and infrared telescopes}}

\begin{document}

\maketitle

\begin{abstract}

The Thirty-Meter Telescope is an ambitious project to build a giant segmented 
mirror telescope with fully integrated adaptive optics systems that will 
produce diffraction-limited images. A powerful suite of instruments is being 
developed that, coupled with the ability to rapidly switch between targets and 
instruments, will allow TMT to take advantage of GRBs to probe fundamental physics
in extreme conditions and as the ultimate tomographic beacons, especially if some 
are as far as z $\sim 10$. This article gives a brief summary of TMT and its instruments,
and some examples of the potential offered by observation of GRBs.

\end{abstract}

\section{Introduction}
The TMT project is currently an equal partnership between Caltech, the University of 
California, and the Association of Canadian Universities for Research in Astronomy 
(ACURA) to construct a 30m telescope. TMT will have a 30-m f/1 primary mirror 
composed of 738 1.2m segments. The final focal ratio of the telescope will be f/15, and 
the field of view will be 20 arcminutes. Sites are being tested in northern Chile, Hawaii 
and Mexico.

The instruments and their associated adaptive-optics (AO) systems will be located on two 
large Nasmyth platforms, and each instrument station will be addressed by the articulated 
tertiary mirror. Although both seeing-limited and AO observing modes will be supported 
at TMT, it is clear that AO will be key in realizing the full scientific potential of the ``D$^4$ 
advantage" offered by such a large aperture.  The telescope, enclosure, AO subsystems 
and instruments are therefore being designed simultaneously as an end-to-end system 
under stringent requirements imposed by AO-based science. 

 Some key system features 
that will benefit GRB observations include:
\begin{itemize}
\item Rapid response: TMT is designed, as a system, to slew and acquire targets, set up 
active and adaptive optic systems, and be ready to observe with any instrument in 
less than 10 minutes.
\item Adaptive optics: TMT's AO systems will deliver high strehl images in the NIR 
and MIR, resulting in a D$^4$ advantage. Laser tomography adaptive optics will 
substantially improve the image quality in the visible. Since GRBs are initially 
point sources, they will benefit the most from full AO correction, unlike distant 
galaxies.
\end{itemize}

\section{Instrumentation}

Most of the proposed instruments will capitalize on the D$^4$ efficiency gain and exquisite 
spatial resolution (7 milliarcsec images in J band) offered by diffraction-limited images. 
TMT instruments will be able to address a broad range of GRB science topics, including:

a) Identification of optical counterparts:
\begin{itemize} 
\item
WFOS (Wide Field Optical Spectrograph): Very efficient imaging and low 
resolution spectroscopy simultaneously in two wavelength bands 0.34 - 0.6$\mu$; 
$0.6 - 1.0\mu$
\item
IRIS: Low spectral resolution (R = 4000) integral field spectroscopy and imaging  
from $0.8 - 2.5\mu$, assisted by high strehl adaptive optics.
\end{itemize}
b) IGM, ISM, Chemical Evolution of the Universe, Fundamental Physics
\begin{itemize}
\item
HROS: Very efficient high (R = 50 - 100, 000) resolution spectroscopy from $0.34 
- 1.0\mu$.  S/N = 100 at m$_{AB}$ $\sim 20$ for R = 50,000
\item
bNIRES: High resolution (R = 50, 000) spectroscopy from $0.8 - 2.5\mu$. Assisted 
by high strehl adaptive optics (D$^4$ advantage). Continuum sensitivity (1hr, $100\sigma$): 
Y, J or H $\sim17.0$. For z = 7, NIRES spectra will cover Ly$\alpha$, Si II, Si IV, C IV, Ni 
II, Al III, Cr II and Zn II.
\item
rNIRES: R = 100,000 3-5$\mu$ spectroscopy, fed by a mid IR AO system or by an 
adaptive secondary. Continuum sensitivity (1hr, 100$\sigma$): L = 13.5, M = 11.5
\end{itemize}

c) Properties of Host Galaxies
\begin{itemize}
\item
IRIS: Integral field spectroscopy with spatial resolutions of better than 100pc for all z $>$ 1. 
In direct imaging, IRIS will reach point sources as faint as K = 28 ($K_{AB} = 30$)  
($3\sigma$) in 3 hours.
\end{itemize}

\section{Summary}

The design of the TMT observatory offers huge potential to exploit the benefits of GRBs. 
More details of TMT and its instruments can be found in \cite{ref:ce, ref:cs}


\begin{thebibliography}{0}

\bibitem{ref:ce} \BY{Crampton D. and Ellerbroek B.}
  in \TITLE{The Scientific Requirements for Extremely Large Telescopes, IAU Symp., 232},
                  edited by \NAME{Whitelock P.A., Dennefeld, M. \atque Leibundgut B.}
                  (Cambridge University Press, Cambridge) 2006, pp.~410-419.
		  
\bibitem{ref:cs} \BY{Crampton D. \atque Simard L.}
  in \TITLE{Ground-based and Airborne Instrumentation for Astronomy},
                  edited by \NAME{McLean I.S. \atque Iye M.}
                  (Proc. SPIE vol 6269) 2006, pp.~62691T1-15
\end{thebibliography}
\end{document}